# Investigation on synthesis and physical properties of metal doped picene solids

Takashi Kambe,[1]* Xuexia He,[2] Yosuke Takahashi,[1] Yusuke Yamanari,[1] Kazuya Teranishi,[2] Hiroki Mitamura,[2], Seiji Shibasaki,[1] Keitaro Tomita,[1] Ritsuko Eguchi,[2] Hidenori Goto,[2,6] Yasuhiro Takabayashi,[2] Takashi Kato,[3] Akihiko Fujiwara,[4] Toshikaze Kariyado,[5] Hideo Aoki,[5] and Yoshihiro Kubozono[2,6*]

[1]Department of Physics, Okayama University, Okayama 700-8530, Japan

[2]Research Laboratory for surface Science, Okayama University, Okayama 700-8530, Japan

[3]Institute for Innovative Science and Technology, Graduate School of Engineering, Nagasaki Institute of Applied Science, Nagasaki 851-0121, Japan

[4]Japan Synchrotron Radiation Research Institute, SPring-8, Hyogo 679-5198, Japan

[5]Department of Physics, The University of Tokyo, Hongo, Tokyo 113-0033, Japan

[6]Research Centre of New Functional Materials for Energy Production, Storage and Transport, Okayama University, Okayama 700-8530, Japan

We report electronic structure and physical properties of metal-doped picene as well as selective synthesis of the phase that exhibits 18 K superconducting transition. First, Raman scattering is used to characterize the number of electrons transferred from the dopants to picene molecules, where a softening of Raman scattering peaks enables us to determine the number of transferred electrons. From this we have identified that three electrons are transferred to each picene molecule in the superconducting doped picene solids. Second, we report pressure dependence of $T_c$ in 7 K and 18 K phases of $K_3$picene. The 7 K phase shows a negative pressure-dependence, while the 18 K phase exhibits a positive pressure-dependence which cannot be understood with a simple phonon mechanism of BCS superconductivity. Third, we report a new synthesis method for superconducting $K_3$picene by a solution process with monomethylamine, $CH_3NH_2$. This method enables us to prepare selectively the $K_3$picene sample exhibiting 18 K superconducting transition. The method for preparing $K_3$picene with $T_c$ = 18 K found here may facilitate clarification of the mechanism of superconductivity.

Corresponding author: Takashi Kambe, kambe@cc.okayama-u.ac.jp & Yoshihiro Kubozono, kubozono@cc.okayama-u.ac.jp

I. Introduction

Recently a new class of organic superconductors has been discovered in aromatic systems. They are solids of hydrocarbons that include picene, coronene, phenanthrene and 1,2:8,9-dibenzopentacene,[1-6] doped with metal atoms. Namely, the superconductivity was first discovered in potassium-doped picene, $K_3$picene, which showed two different superconducting transition temperatures, one with $T_c$ = 7 K and the other as high as 18 K.[1] This has been followed by other studies, and the highest $T_c$ among these hydrocarbon superconductors to date attains 33 K observed in $K_{3.45}$dibenzopentacene,[6] whose $T_c$ is much higher than the highest $T_c$ (14.2 K at 8.2 GPa[7] in $\beta'$- (BEDT-TTF)$_2$ICl$_2$) in charge-transfer organic superconductors. Thus the hydrocarbon superconductors are very attractive from viewpoints of development of new high-$T_c$ superconductors as well as fundamental physics of superconductivity. Theoretical calculations for picene superconductors were also achieved, which suggests that the electron-phonon coupling is strong,[8,9] the conduction band comprises four bands arising from two LUMO orbitals,[10] and that strong hybridization between the dopants and molecules invalidates a rigid-band picture.[10]

The departure from the rigid-band picture was experimentally evidenced by photoemission spectroscopy.[11] This photoemission study clearly showed a metallic ground state for potassium-doped picene films. Our recent resistivity data also indicate a metallic behavior for the $K_3$picene phase.[12] Further, a Pauli paramagnetic susceptibility was observed for a $K_3$picene bulk sample.[1] These results support a metallic ground state for $K_3$picene.

The $T_c$ for the solid $K_3$picene was found to be either 7 or 18 K,[1,2] while the $T_c$ of $K_3$phenanthrene was reported to be ~5 K.[3] Donation of three electrons to an aromatic hydrocarbon molecule was reported to be a key for superconductivity.[1-6] Very recently, specific heat has also been measured for $Ba_{1.5}$phenathrene,[13,14] which suggests an s-wave, single superconducting gap and an intermediate electron-phonon coupling strength. While the nominal $x$ value (number of metal atoms intercalated) in the superconducting samples of A$_x$hydrocarbon, AE$_x$hydrocarbon and Ln$_x$hydrocarbon (A: alkaline-metal atom, AE: alkaline-earth metal atom, Ln: lanthanide atom) are $x$ = 3, 1.5 and 1, respectively, it is imperative to determine the real $x$ value. Rietveld refinements for X-ray powder diffraction patterns have been extensively used for identifying $x$ in metal-doped $C_{60}$ compounds.[15-17] Since single crystals cannot simply be obtained in the intercalation compounds, the Rietveld refinement is valuable. For the hydrocarbon superconductors, however, the Rietveld refinements have never been achieved because of its low crystal symmetry where the doped metal atoms do not occupy special symmetric sites unlike in the doped $C_{60}$ compounds.

On the other hand, Raman scattering is known to be a powerful tool for determining the $x$ value in $A_xC_{60}$.[18-20] The $A_g$ peak, which is observed at 1469 cm$^{-1}$ for pure $C_{60}$ (with $I_h$ symmetry), shifts to lower frequencies when alkaline metal atoms are doped into $C_{60}$ solids. Actually, the peak shifts by 6 – 7 cm$^{-1}$ for one electron donation to a $C_{60}$ molecule, namely, the $A_g$ peak is observed at 1452 cm$^{-1}$ for $K_3C_{60}$,[18] 1448 cm$^{-1}$ for $Rb_3C_{60}$[19] and 1447 cm$^{-1}$ for $Cs_3C_{60}$,[20] providing a good probe for determining the amount of doping. If we now turn to hydrocarbon systems, we previously suggested that some of the peaks shift to lower frequencies with increasing the $x$ value in $K_x$picene[2] in a similar manner as in $A_xC_{60}$,[18-20] which indicates that the number of electrons on picene can be determined from the shift of Raman peak from that in pure picene. The number of electrons on phenanthrene molecule in $A_x$phenanthrene was successfully determined from the Raman spectrum, which indicates that the superconducting component is $A_3$phenanthrene. Thus three electron donation seems to be a key for superconductivity,[3-5] but we have definitely to elaborate this. So the first purpose of the present study is to systematically investigate the Raman scattering for $A_x$picene (A: K and Rb) to correlate the number of electrons on the hydrocarbon molecules with the Raman frequency. The experimental Raman frequencies have been measured for a wide range of $x = 0 – 5$ in $A_x$picene, which is compared with theoretical frequencies for the Raman peaks. From this, we have evaluated the number of electrons per picene molecule to clearly determine the $x$ value in the superconducting phases.

Now, in characterizing physical properties of picene superconductors, it is intriguing to compare the picene superconductors with more familiar carbon-based superconductors, *i.e.*, $C_{60}$ and graphite compounds intercalated with alkaline or alkaline earth metal atoms, where the highest $T_c$ to date is 33 K for $RbCs_2C_{60}$ among $C_{60}$ compounds,[21] and 11.5 K for $CaC_6$ among graphite compounds.[22] One probe of these materials is the pressure dependence of $T_c$: indeed, $C_{60}$ and graphite superconductors have opposite tendencies; $K_3C_{60}$ ($T_c$ = 18 K) shows a large negative pressure coefficient ($dT_c/dP$ = -7.8 K GPa$^{-1}$),[23] while $CaC_6$ has a positive one ($dT_c/dP$ = 0.42 - 0.48 K GPa$^{-1}$).[24] In the previous paper, we reported briefly the pressure dependence of $T_c$ in 7 K phase of $K_3$picene,[2] but the detailed analysis was not presented. Very recently, positive pressure dependence of $T_c$ in $A_x$phenanthrene, $AE_x$phenanthrene and $Ln_x$phenanthrene was reported,[3-5] which cannot be understood with a simple phonon mechanism of BCS superconductivity. Therefore it is intriguing to examine the pressure dependence of $T_c$ in the $T_c$ = 7 and 18 K phases of picene superconductors, which may provide an important key for investigating the mechanism of superconductivity. Thus the second purpose of the present paper is to explore the pressure dependence of 7 K and 18 K phases of $K_3$picene, respectively, here reported in a pressure region of 1 bar to 1.2 GPa. We shall show that 18 K phase of $K_3$picene has a positive pressure dependence of $T_c$, while 7 K phase has a negative one. The pressure dependence is discussed from the viewpoints of crystal and electronic structures of picene superconductors.

The third purpose of the present paper is on the fabrication method for $K_3$picene samples. A still unclear question is why we have two phases with $T_c$ = 7 and 18 K. The usual annealing method (or solid-reaction method) does not allow us to produce selectively 7 or 18 K phases. Even a precise control of nominal composition in $K_x$picene from $x$ = 2.6 to 3.3 produced both 7 K and 18 K superconductors.[1,2] Here we have prepared $K_3$picene superconductor by a solution process with monomethylamine, $CH_3NH_2$, as a solvent, in search for a synthetic method for selectively preparing 7 K or 18 K phase. The solution process has previously been used to control polymorphs of $Cs_3C_{60}$, which is a pressure-induced superconductor ($T_c$ = 38 K (A15 phase) and $T_c$ = 35 K (face-centered cubic phase) at 15 kbar).[25,26] Here we shall show that the solution process with $CH_3NH_2$ for $K_3$picene enables us to prepare selectively the $K_3$picene sample exhibiting only $T_c$ = 18 K.

**II. Experimental**

Picene (purity: 99.9%) was purchased from NARD Co Ltd., which was used for the experiments without further purification. Alkaline or alkaline earth metal was mixed with picene powder at nominal $x$ value in a pyrex glass tube to fabricate $A_x$picene. In the annealing method, the pyrex tube was pumped and sealed at $10^{-5}$ Torr. The tubes were annealed at temperatures as high as ~443 K for ~10 days. The samples obtained after annealing were black in color. The Raman spectrum and magnetic susceptibility were measured for the samples treated in a glove box without any exposure to air. The Raman scattering was measured at 295 K with Raman spectrometer (JASCO NRS-3100) with an excitation energy having a wavelength $\lambda =$ 785 nm, while magnetization, $M$, was measured with a SQUID magnetometer (Quantum Design MPMS2) in the temperature region > 2 K under ambient and high pressures. A magnetic field, $H$, of 20 Oe was applied in measuring $M/H$. For the measurement of pressure dependence of superconductivity, a Cu-Be piston-cylinder type pressure-cell was used. The hydrostatic pressure was mediated by Daphne oil. The applied pressure region was from 1 bar to 1.2 GPa. A small piece of crystalline Sn or Pb was put inside the pressure cell along with the sample to monitor the exact pressure.

In the solution method for preparing $A_x$picene, on the other hand, $CH_3NH_2$ was introduced from the $CH_3NH_2$ gas bottle into the glass vessel containing picene powder and K metal at nominal stoichiometric amounts. The picene and K were completely dissolved in $CH_3NH_2$ by stirring at 223 K for 3 h. The color of solution was dark-green. After complete dissolution, liquid $CH_3NH_2$ was removed at 300 K until the base pressure reaches $10^{-4}$ Torr. The black powder sample obtained after removing $CH_3NH_2$ was moved into quartz tube in an Ar filled glove box. The tube, dynamically pumped and sealed at $10^{-6}$ Torr, was annealed at 443 K for 70 h to remove $CH_3NH_2$ completely. Special care was taken in the annealing procedure because high temperature annealing above 490 K for removal of $CH_3NH_2$ produces a very toxic material, KCN; actually even below 490 K, KCN may be produced. Powder X-ray diffractions for $K_x$picene were measured by RIGAKU RINT-TTR III and with synchrotron radiation (KEK-PF, SPring-8 and ESRF). The unit-cell parameters for $K_x$picene were determined by the LeBail analysis program in GSAS package.[27] Theoretical calculation of Raman frequencies and intensities for $A_1$, $A_2$, $B_1$ and $B_2$ modes for neutral picene and its anions, picene$^{y-}$ ($y = 0 - 5$), with optimized structure ($C_{2v}$ symmetry) was performed with B3LYP program based on hybrid Hartree-Fock (HF) / density functional theory (DFT) with the 6-31G* basis set.[28-30]

**III. Results and Discussions**

**III-1. Characterization of the number of doped metal atoms in $K_x$picene and $Rb_x$picene**

Figures 1(a) and 2(a) show Raman spectra for samples of $K_x$picene and $Rb_x$picene ($x = 0 - 5$), respectively, at $500 - 1800$ cm$^{-1}$. A pronounced peak at 1378 cm$^{-1}$ marked with an asterisk for pristine picene can be assigned to a superposition of $\nu_{20}$ and $\nu_{21}$ $A_1$ vibration modes of picene molecule, where $\nu_n$ stands for the $n$th $A_1$ vibration mode from the bottom. The $\nu_{20}$ and $\nu_{21}$ modes in picene$^{y-}$ ($y = 0 - 5$) are schematically depicted in Fig. 3. These vibration modes are suggested to provide strong electron-phonon coupling.[8] In Figs. 1(b) and 2(b) the theoretical frequencies and intensities of Raman-active $A_1$, $A_2$, $B_1$ and $B_2$ vibration modes are shown along with the experimental Raman spectra, where the experimental peaks are seen to agree well with the theoretical results. Figures 1 and 2 also show that the pronounced peak observed at 1378 cm$^{-1}$ for pristine picene shifts to lower frequencies with an increase in $x$ for both $K_x$picene and $Rb_x$picene samples.

The average values of $\nu_{20}$ and $\nu_{21}$ calculated theoretically are plotted for each picene molecule in Figs. 4(a) and (b). The experimental frequencies for the pronounced peak, which are ascribable to superposition of $\nu_{20}$ and $\nu_{21}$, are plotted as a function of $x$ for $K_x$picene and $Rb_x$picene in Figs. 4(a) and (b), respectively. As seen from Fig. 4(a), the experimental frequencies in all the $K_x$picene samples basically fall upon three discrete values, 1378, 1344 and 1313 cm$^{-1}$, which are consistent with those predicted theoretically for picene, picene$^{2-}$ and picene$^{3-}$ (respective dashed lines). This implies that only two phases, $K_2$picene and $K_3$picene, can be produced in doped $K_x$picene samples. As for $K_1$picene, this separates into two phases, picene and $K_2$picene, as seen from Figs. 1(b) and 4(a). The same plots suggest that $K_{1.5}$picene and $K_2$picene samples decompose into three phases, picene, $K_2$picene and $K_3$picene.

Further, a Raman peak was observed in $K_3$picene samples at 1328 cm$^{-1}$ in addition to those for $K_2$picene (at 1344 cm$^{-1}$) and $K_3$picene (1313 cm$^{-1}$), where the value (1328 cm$^{-1}$) is intermediate between the latter two. Thus the Raman peak at 1328 cm$^{-1}$ may imply the existence of a phase with a fractional $x$ ($K_{2.5}$picene), although the phase is not a main one. However, two Raman peaks should be observed at 1344 and 1313 cm$^{-1}$ because Raman scattering reflects vibration of molecule even if the $K_{2.5}$picene phase appears as statistically averaged structure in this sample. Thus the origin of the Raman peak at 1328 cm$^{-1}$ remains to be clarified. If a dynamic conversion of picene$^{2-}$ and picene$^{3-}$ produces the peak at 1328 cm$^{-1}$, the Raman scattering should be observed at low temperatures with the single peak splitting into peaks at 1344 and 1313 cm$^{-1}$. When nominal $x$ value was increased above 3, only $K_2$picene and $K_3$picene were observed, while $K_4$picene and $K_5$picene phases could not be fabricated, which indicates that the maximum $x$ value is three. From these plots, we conclude that only two phases, $K_2$picene and $K_3$picene, can be realized by an intercalation of K atoms into picene samples, while $K_1$picene is unstable. In Fig. 4(a), the plots for phases with the same $x$ value as the nominal $x$ in the $K_x$picene sample is indicated in red (e.g., the Raman peak assigned to $K_3$picene phase in the $K_3$picene sample).

Here, we should stress that the prepared individual $K_x$picene sample does not always contain all the crystal phases described above, but that some solid samples contain only one or two crystal phases. The number of times the phases appeared is given in parentheses in Fig. 4(a). The numbers then indicate the frequency with which the crystal phases appear. For example, all $K_3$picene samples (19 of them) contained the $K_3$picene crystal-phase, and 58% in 19 $K_3$picene samples showed only a peak ascribable to $K_3$picene crystal phase *i.e.*, a single phase of $K_3$picene. $K_2$picene samples produce a single phase of $K_2$picene (20%), picene+$K_2$picene+$K_3$picene phases (20%), a single phase of $K_3$picene (20%) and $K_2$picene+$K_3$picene (40%). $K_1$picene samples resulted in a single phase of $K_2$picene (25%) and picene+$K_2$picene phases (75%). $K_{1.5}$picene samples provided

picene+$K_2$picene phases (20%), $K_2$picene+$K_3$picene (20%) and picene+$K_2$picene+$K_3$picene phases (40%) in which each fraction of phase (peak intensity) was almost the same. The $K_{1.5}$picene samples provided only a single phase of $K_2$picene (20%). It is suggested from these results that K metal in solid picene may not completely react in most of the samples except for those samples producing a single phase. Judging from the probability, 58%, of realization of the single phase, it is suggested that $K_3$picene is more stable than $K_2$picene (25% for a single phase). Therefore, it is of interest to investigate theoretically the energetic stability of $K_x$picene phase.

As seen from Figs. 2(b) and 4(b), the $\nu_{20}$, $\nu_{21}$ Raman frequencies in $Rb_x$picene samples also suggest existence of only three phases (picene, $Rb_2$picene and $Rb_3$picene) as in $K_x$picene. The observed frequencies, 1378, 1345 and 1313 cm$^{-1}$, can be assigned to picene, picene$^{2-}$ and picene$^{3-}$, respectively; the values are consistent with those of $K_x$picene. It can be concluded from the plots (Fig. 4(b)) that $Rb_1$picene sample provides only two phases, picene and $Rb_2$picene, while $Rb_2$picene and $Rb_3$picene solid samples provide only three phases (picene, $Rb_2$picene and $Rb_3$picene). Thus, increasing nominal $x$ leads to a realization of phases with larger integer $x$ values, in the same manner as in $K_x$picene (Fig. 4(a)). Increasing nominal $x$ above 3 produces $Rb_3$picene and picene as well as a new phase of $Rb_{2.5}$picene exhibiting a Raman frequency at 1323 cm$^{-1}$, which was also found in $K_x$picene. In the same manner as $K_x$picene, the stability of $Rb_3$picene is confirmed because of high probability, 44%, of formation of a single phase in the $Rb_3$picene samples.

The magnetic susceptibility has been measured for all the samples of $K_x$picene and $Rb_x$picene. The $M/H - T$ curves for $K_x$picene samples that show the existence of $K_3$picene phase (with the Raman peak at ~1313 cm$^{-1}$) exhibits a clear drop at 7 or 18 K, indicating superconducting transitions. Conversely, all the samples that have no $K_3$picene phase exhibit no such behaviors. These results clearly indicate that superconducting phase in $K_x$picene relates closely to $K_3$picene phase. The $M/H$ of $Rb_x$picene samples also show the same behavior, *i.e.*, the samples showing the existence of $Rb_3$picene phase have the superconducting transition, while the $Rb_3$picene samples that do not exhibit the Raman peak at ~1313 cm$^{-1}$ exhibit no superconducting transitions, indicating that superconducting phase relates to Rb$_3$picene. Thus, we have identified that superconducting phases in K$_x$picene and Rb$_x$picene can be closely associated with K$_3$picene and Rb$_3$picene crystal phases, respectively, from the *M/H* and Raman measurements.

**III-2. Pressure dependence of superconducting transition temperature in 7 K and 18 K phases**

We now turn to the pressure effect. We first display the *M/H* – *T* curves at various pressures for the 7 K phase of K$_3$picene. Figure 5 shows that the $T_c$ shifts downward with pressure, but only gradually. The midpoint of the superconducting transition, $T_c^{mid}$, shown in Fig. 5(b), decreases linearly with pressure up to 1 GPa, with $dT_c^{mid}/dP$ = -0.3 K GPa$^{-1}$, which can be determined unambiguously because of the absence of any inhomogeneous broadening of the superconductive transition with increasing pressure. So $dT_c/dP$ of the 7 K phase is negative as in K$_3$C$_{60}$, but the coefficient is an order of magnitude smaller. K$_3$C$_{60}$ has a three-dimensional (3D) electronic band structure, so that pressure is expected to increase the bandwidth with a decreased density of states at the Fermi energy, $N(\varepsilon_F)$,[23] which in turn decreases $T_c$, regardless of a softening of some C$_{60}$ vibration modes.[31] The intercalated graphite, CaC$_6$, on the other hand, has a positive pressure dependence, where the electronic structure around the Fermi energy is shown to contain a large component of the so-called interlayer states, whose amplitudes reside in between the graphene layers, conferring a 3D character on the electronic structure.[24,32] In this compound, while the $N(\varepsilon_F)$ decreases with pressure, a large softening of an in-plane Ca vibration mode under pressure is shown to cause an increase in the electron-phonon coupling $\lambda$, which is considered to overcome the reduction of $N(\varepsilon_F)$, leading to an enhanced $T_c$.

Now we turn to the pressure dependence of $T_c$ in the 18 K phase of K$_3$picene. Figure 6(a) shows the *M/H* against *T*, normalized by the value at 50 K for clarity, for various pressures in the 18 K superconducting phase. We can immediately notice that the $T_c^{onset}$, shifts *upward*: The $T_c^{onset}$, shown in Fig. 6(b) against pressure, increases linearly with pressure up to 1.2 GPa, with $dT_c^{onset}/dP$ = 12.5 K GPa$^{-1}$. The large positive pressure dependence found here contrasts sharply with the negative pressure dependence in the 7 K phase. The positive pressure dependence has also observed in A$_3$phenanthrene, AE$_{1.5}$phenanthrene and Ln$_1$phenanthrene.[3-5] However, the size of the coefficient, d$T_c^{onset}$/d$P$, is an order of magnitude larger than that, ~1 K GPa$^{-1}$, of phenanthrene superconductor. Since a reduced volume usually implies a smaller density of states at $\varepsilon_F$, this behavior cannot be understood with a simple phonon mechanism of BCS superconductivity.

In K$_3$picene, the electronic structure calculation[10] indicates that the interlayer band lies well above the Fermi energy, while the character of the conduction band primarily originates from the LUMO/LUMO+1 orbitals of the picene molecule. More importantly, the wave functions in the conduction band have large amplitudes on the doped K atoms, so that the dopants act not only as a source of charge transfer, but dopant orbits significantly hybridize with the aromatic molecular orbits. Through this mechanism, the coupling between the molecules and dopants should be strong. As reported previously,[2] when the $x$ value in K$_x$picene and Rb$_x$picene is increased, the $a$-axis is expanded while the $b$- and $c$-axes are shrunk, leading to a reduction of unit-cell volume. This implies that the picene molecules become more densely packed when doped with K. This peculiarity comes from a deformation of the in-layer, herringbone arrangement of picene molecules.[1,2] A theoretical structure optimization[10] indeed indicates that the angle between picene molecules in the herringbone arrangement changes dramatically from 61° to 114° as the metal atoms are doped. These are further indications that the picene molecules and metal atoms form a strongly-knit layer. Thus we expect, first of all, that this will cause the difference in the pressure effect between 7 K and 18 K phases, given the difference in structural phases as described later. The strong coupling between molecules and dopants may also account for the small pressure dependence of $T_c$ in the 7 K phase. Here it should be noticed that $T_c$ (= 7 K) is invariant in A$_3$picene for A = K and Rb,[1] unlike in fulleride superconductors.[33] The small chemical-pressure effect may also be explained well by the strong coupling described above.

**III-3. Selective preparation of 18 K superconducting phase in K$_3$picene**

Figure 7(a) shows the $M/H - T$ curve in zero-field cooling (ZFC) for K$_{3.1}$picene sample (nominal $x$ = 3.1) that was prepared by the solution process. This data was reported and briefly discussed in Ref. 2. Before annealing the sample or removing $CH_3NH_2$ from the precursor, the $M/H$ shows a Pauli-like, temperature-independent behavior with a weak increase below 10 K. After the sample is annealed at 443 K for 70 h, the $M/H$ begins to show an abrupt decrease with the $T_c^{onset}$ = 18 K and the $T_c^{mid}$ = 17 K. This superconducting transition coincides with that for the 18 K phase of $K_3$picene superconductor prepared by solid-reaction method, or annealing method.[1,2] The maximum shielding fraction is still as low as 0.1%, a value lower by an order of magnitude than that, 1.2 %, for 18 K superconductor prepared by annealing method.[1] However, we notice that the solution-reaction method can produce 18 K superconducting phase more effectively than the solid-state method. In fact all the $K_3$picene samples prepared by the solution method show a clear decrease in $M/H$ at 18 K. The nominal $x$ value for producing the superconducting phase with the solution process is confined to 2.9 – 3.1, which again suggests that three K atoms per picene molecule is a key doping level.

Figure 7(b) shows Raman scattering spectra for the $K_3$picene samples prepared by the solution method. One sample (curve A in Fig. 7(b)) was annealed at 443 K for 70 h in vacuum to remove $CH_3NH_2$, while the other (curve B) was not annealed. For comparison, the spectra for the $K_3$picene sample prepared by the solid-reaction method (curve C) as well as for pristine picene (curve D) are also shown. We can see that the Raman spectrum for the $K_{3.1}$picene sample prepared by the solution method (A) coincides with that for the $K_{3.0}$picene prepared by the solid-reaction method (C), where the peak shifts to a lower frequency by 67 $cm^{-1}$ from 1378 $cm^{-1}$ for pristine picene, which suggests that these samples possess the same number of electrons per picene molecule, *i.e.*, these can be represented as $K_3$picene. This result indicates the effectiveness of annealing at 443 K in vacuum for the removal of $CH_3NH_2$, and that $CH_3NH_2$ molecules are scarcely left in $K_{3.1}$picene sample (A). On the other hand, the peak for the $(CH_3NH_2)K_{3.0}$picene sample (B), which was prepared by solution method without annealing, shifts to a lower frequency by 48 $cm^{-1}$, showing that 2.5 electrons are transferred to picene molecule from K atoms. This implies that the remaining $CH_3NH_2$ molecule may capture 0.5 electrons from a picene molecule, *i.e.*, a back-electron transfer from picene molecule.

Figures 8(a) and (b) show X-ray diffraction patterns for the $(CH_3NH_2)_xK_3$picene and $K_3$picene samples, respectively. The former was prepared by the solution method without annealing, while the latter was prepared by the solution method with annealing at 443 K for 70 h. The lattice parameters for the two samples obtained from the LeBail fit to X-ray diffraction patterns (Fig. 8) are listed in table 1; the space group was assumed to be $P2_1$ in these samples, in the same manner as picene and $K_3$picene prepared by solid reaction method.[1,2] We can see from this table that the lattice parameters, $a$ and $c$, for the $K_3$picene sample prepared by the solution method without annealing increase from those in pristine picene. Specifically, the $c$ axis expands markedly, suggesting that $CH_3NH_2$ molecules are mainly intercalated into the space between the herringbone (*ab*-plane) layers of picene molecules.

The values of $a$ and $c$ in the $K_3$picene sample, prepared by the solution method with annealing, are smaller than those in the sample without annealing, which indicates that $CH_3NH_2$ was basically removed in the sample annealed at 443 K in vacuum, as consistent with the results obtained from the Raman scattering. The $K_3$picene sample prepared by the solid reaction (C in Fig. 7(b)) showed $T_c$ of 7 K, while the $K_3$picene sample (A in Fig. 7(b)) prepared by the solution method with annealing showed $T_c$ of 18 K. All the lattice parameters in the $K_3$picene sample prepared by the solution method with annealing are larger than those in pure picene, especially the $c$ expands, suggesting that the crystal structure of $K_3$picene phase prepared by solution method is different from that (7 K phase) prepared by the solid reaction, *i.e.*, the K atoms in $K_3$picene prepared by the solution method may be intercalated into the space between *ab*-layers.

In fact, the theoretical calculation[10,34,35] (as summarized in table 2 for the structure parameters) shows an existence of two kinds of doping structures: (i) the one (denoted as $K_3$picene) with the dopants inserted within the herringbone-arranged picene layer (*ab*-layer), and (ii) another with some dopants intercalated in the interlayer regions as well, where the latter is meta-stable but does exhibit a local energy minimum. The structure is denoted as $K_2K_1$picene where two K atoms are inserted into the *ab*-layer while one K atom is intercalated into the space between *ab*-layers.[2,10] The lattice constants determined by X-ray diffraction indicate the expansion of *a*-axis and shrinkage of *c*-axis in the 7 K phase prepared by solid reaction, suggesting the intercalation of K atoms into the picene layer (*ab*-layer), which is consistent with the location of K atoms with intralayer insertion obtained theoretically.[10] On the other hand, some of the K atoms in K$_3$picene prepared by solution method may be intercalated into the space between *ab*-layers, which is consistent with the theoretical prediction that the structure with both intralayer and interlayer dopants is a meta-stable structure.[10,34,35] Thus the theoretical structure (K$_2$K$_1$picene with two intra-layer and one inter-layer picene molecules) may possibly be related to the 18 K phase prepared by solution method, but this definitely requires further confirmation.

Superconducting mechanism and the symmetry of the gap function have yet to be known. The previous theoretical calculation suggests that the electron-phonon coupling is sufficiently strong to roughly account the size of $T_c$ in doped picene.[8,9,36] At the same time, however, a possible relevance of the strong electron-electron correlation is pointed out in other theoretical works.[37-39] The multiple structures, found here to be dependent on the sample fabrication method, is consistent with the previous theoretical suggestion that there are multiple meta-stable structures in the doped picene.[10,34,35] More importantly, if the identification of the 18 K phase of the superconductivity to be the structure with the intralayer plus interlayer insertion of dopants is correct, this implies the superconductivity is dominated by details in the doping structure, and this may give a crucial clue in exploring the superconductivity mechanism, including some unconventional ones. Indeed, here we find that the 7 K and 18 K phases react in opposite manners against the applied pressure, which may provide an important test in sorting the mechanism. The first thing we should check theoretically is how the density of states (DOS) changes against pressure. As shown theoretically[10,34,35], the conduction band comprises multiple orbitals, so that the DOS may show a nontrivial dependence on the pressure. The next interest is how the shape of the Fermi surface, which can be crucial in unconventional superconductivity, changes against pressure. This is interesting, since even at ambient pressure the Fermi surface is shown to be a composite of pieces with different dimentionalities[10,34,35]. Such a study is under way, and will be reported elsewhere.

## IV. Summary and discussion

The conclusion of the present paper is three-fold: (i) We have performed a characterization of the number of electrons on picene molecule by use of Raman scattering, and two different phases of $A_2$picene and $A_3$picene were found in the $A_x$picene samples, indicating the absence of $A_1$picene phase. From the Raman scattering of superconducting $A_x$picene samples, it has been found that the $A_3$picene phase relates closely to superconducting phase. (ii) The pressure study revealed that the 18 K phase has a positive $dT_c^{mid}/dP$ as in superconducting phenanthrene[3-5], while the 7 K phase has a negative coefficient. The latter is understandable from the BCS picture, while the former does not fit with a simple phonon-mechanism BCS superconductivity. (iii) The preparation of $K_3$picene by a solution method led selectively to 18 K superconducting phase, which is different from the preparation by annealing method which produces 7 or 18 K phase. The $K_3$picene sample prepared by the solution method showed longer $c$ axis than in $K_3$picene with the annealing method, indicating that the K atoms may be intercalated into the space between the herringbone ($ab$) layers.

As for the metallicity of the system, Refs. 40,41 report photoemission results for a vanishing density of states at the Fermi energy, suggestive of an insulating state of the doped picene. However, it has also been found that the photoemission spectrum for pristine picene films significantly varies according to the substrate used,[40] which suggests that the growth orientation of molecules and/or the crystal structure may depend on the substrate, and this may account for the difference in the electronic states (metallic or insulating) in doped picene. Theoretically, Ref. 41 suggests that K-doped picene systems are insulating for all the stoichiometric filling $n$ on the basis of DFT + DMFT calculation. One possible explanation for this discrepancy, apart from the possible difference in the film structure mentioned above, is that the Hubbard $U$ = 1.6 eV[41], extracted from a molecular based calculation[37], is considerably larger than $U \sim 0.8$ eV estimated with the constrained random phase approximation [39]. The precise value of $U$ may be important, since, according to Ref. 41 the metal-insulator transition takes place around $U$ = 0.6-0.8 for $K_3$picene. If the actual $U$ value is on the Mott insulating side, the observed metallicity may come from a small deviation of the filling *n* from an integer value.

Thus, with the present study opening up a new synthesis root for producing the crystal phase (K$_3$picene) exhibiting the 18 K superconducting transition, the correlation of the electronic structure with the arrangement of molecules and doping level should be an interesting future problem.

**Acknowledgment.** The authors appreciate to Takayoshi Yokoya and Hiroyuki Okazaki for valuable discussions on the photoemission study, Masaki Mifune of Department of Pharmaceutical Science, Okayama University, for his valuable assistance in the Raman measurement. HA is benefitted from discussions with Taichi Kosugi of AIST. This study is partly supported by Grant-in-aid (23340104, 23684028, 22244045, 24654105) of MEXT, by the LEMSUPER project (JST-EU Superconductor Project) in Japan Science and Technology Agency (JST), and by Special Project of Okayama University / MEXT. The X-ray diffractions with synchrotron radiation were done under the proposals of KEK-PF (2011G109), SPring-8 (2011A1938) and ESRF (HS-4556).

Table 1. Experimentally obtained lattice parameters for picene and K-doped picene.

| | $a$ (Å) | $b$ (Å) | $c$ (Å) | $\beta$ (°) | $V$ (Å$^3$) |
|---|---|---|---|---|---|
| Pristine picene[a] | 8.472(2) | 6.170(2) | 13.538(7) | 90.81(4) | 708 |
| K$_3$picene[b] | 8.707(7) | 5.912(4) | 12.97(1) | 92.77(5) | 667 |
| (CH$_3$NH$_2$)$_z$K$_3$picene[c] | 8.927(5) | 6.151(1) | 14.476(4) | 94.16(3) | 793 |
| K$_3$picene[d] | 8.571(5) | 6.270(2) | 14.001(3) | 91.68(3) | 752 |

a) Taken from ref. 1.
b) Taken from ref. 1, where the sample was prepared by solid reaction method.
c) Sample prepared by solution method without annealing.
d) Sample prepared by solution method with annealing.

Table 2. Theoretically obtained lattice parameters for K-doped picene.[34] K$_3$picene stands for a structure where all the three K atoms are inserted within the picene layer (with two possible structures A and B), while K$_2$K$_1$picene a structure with two intralayer atoms and one interlayer one.

| | $a$ (Å) | $b$ (Å) | $c$ (Å) | $\beta$ (°) | $V$ (Å$^3$) |
|---|---|---|---|---|---|
| K$_2$K$_1$picene | 8.776 | 6.394 | 13.346 | 94.03 | 747.069 |
| K$_3$picene (A) | 7.421 | 7.213 | 14.028 | 104.53 | 726.848 |
| K$_3$picene (B) | 7.408 | 7.223 | 14.116 | 105.93 | 726.328 |

Figure captions

Fig. 1. Raman scattering spectra in $K_x$picene ($x = 0 - 5$) at (a) 500 – 1800 cm$^{-1}$ and (b) 1250 – 1400 cm$^{-1}$. Raman peaks calculated theoretically are indicated with bars in (b) with $A_1$ ($B_2$) mode in red (blue). The height of the red bars represents the relative intensities of the Raman peak. All the theoretical results are shifted downward by 27 cm$^{-1}$ so that the theoretical average value of $\nu_{20}$ and $\nu_{21}$ $A_1$ modes in neutral picene fits to its experimental average value, 1378 cm$^{-1}$. The arrows indicate averaged values of theoretical $\nu_{20}$ and $\nu_{21}$ $A_1$ modes. The explanation of notation, $\nu_n$, is described in text.

Fig. 2. A similar plot as in Fig.1 for $Rb_x$picene ($x = 0 - 5$).

Fig. 3. $\nu_{20}$ and $\nu_{21}$ $A_1$ modes schematically shown for picene$^{y-}$ ($y = 0 - 5$).

Fig.4. Plots of frequency of experimental and theoretical Raman peaks against the nominal value, $x$, in (a) $K_x$picene and (b) $Rb_x$picene ($x = 0 - 5$). Frequency of the theoretical Raman peak corresponds to the average value of frequencies for $\nu_{20}$ and $\nu_{21}$ $A_1$ modes. The theoretical plots are shifted by 27 cm$^{-1}$ so that the theoretical average value of $\nu_{20}$ and $\nu_{21}$ $A_1$ modes in neutral picene fits to its experimental average value, 1378 cm$^{-1}$. The experimental plots in red refer to the $K_x$picene ($Rb_x$picene) phase produced in the $K_x$picene ($Rb_x$picene) sample (see text). The numerical value in parenthesis stands for the number of times the phase appeared in each sample.

Fig. 5. (a) $M/H - T$ curves for various pressures, and (b) pressure dependence of $T_c^{mid}$ in the 7 K superconducting phase of $K_{3.3}$picene. $dT_c^{mid}/dP$ is determined by a linear fitting in $T_c^{mid} - P$ plot.

Fig. 6. Similar plot as in Fig.5 for the 18 K superconducting phase of $K_{3.3}$picene.

Fig. 7 (a) $M/H - T$ curve in zero-field cooling (ZFC) for $(CH_3NH_2)K_3$picene and $K_{3.1}$picene samples, which were prepared by the solution process (see text). This data was reported and briefly discussed in ref. 2. (b) Raman scattering spectra for the $K_3$picene samples which were prepared by the solution method (curve A), $(CH_3NH_2)K_3$picene (curve B), $K_3$picene sample prepared by the solid-reaction method (curve C), and pristine picene (curve D).

Fig. 8. X-ray diffraction patterns for (a) $(CH_3NH_2)K_3$picene and (b) $K_3$picene samples. Red circles (blue lines) represent the observed (calculated) diffraction patterns. The green line shows the difference between the observed and calculated patterns. Black vertical bars correspond to the predicted peak positions. The lattice parameters for two samples, $(CH_3NH_2)K_3$picene and $K_3$picene, obtained from the LeBail fit to X-ray diffraction patterns are listed in Table 1.

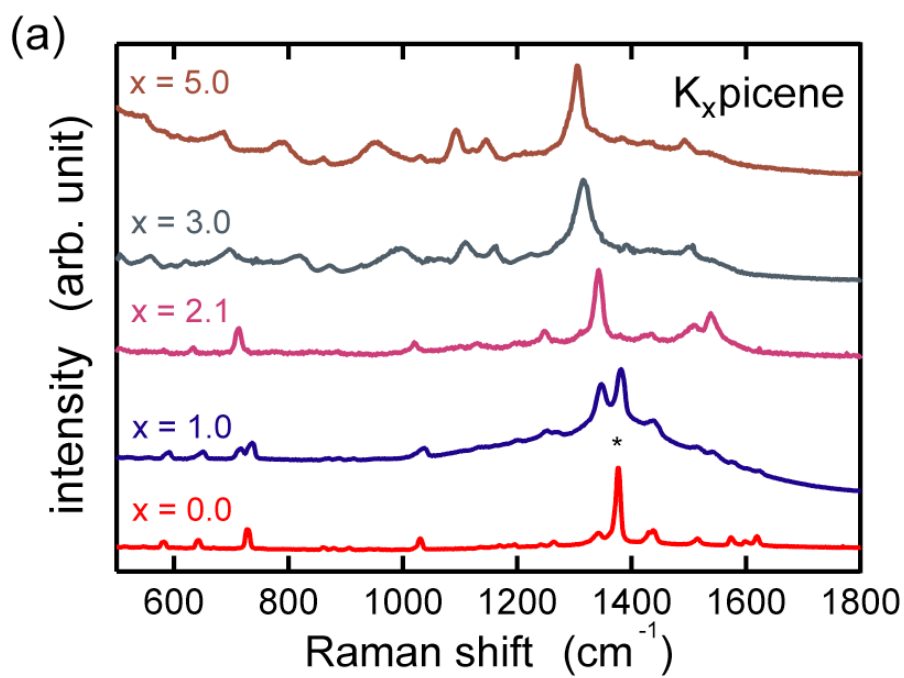

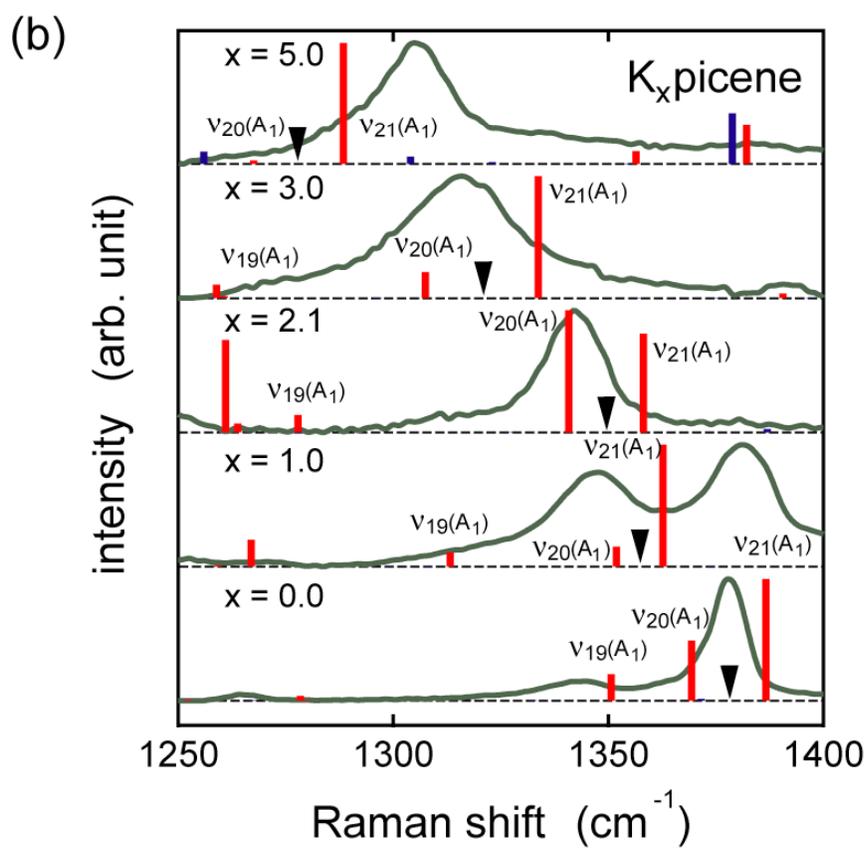

Figure 1. T. Kambe et al.,

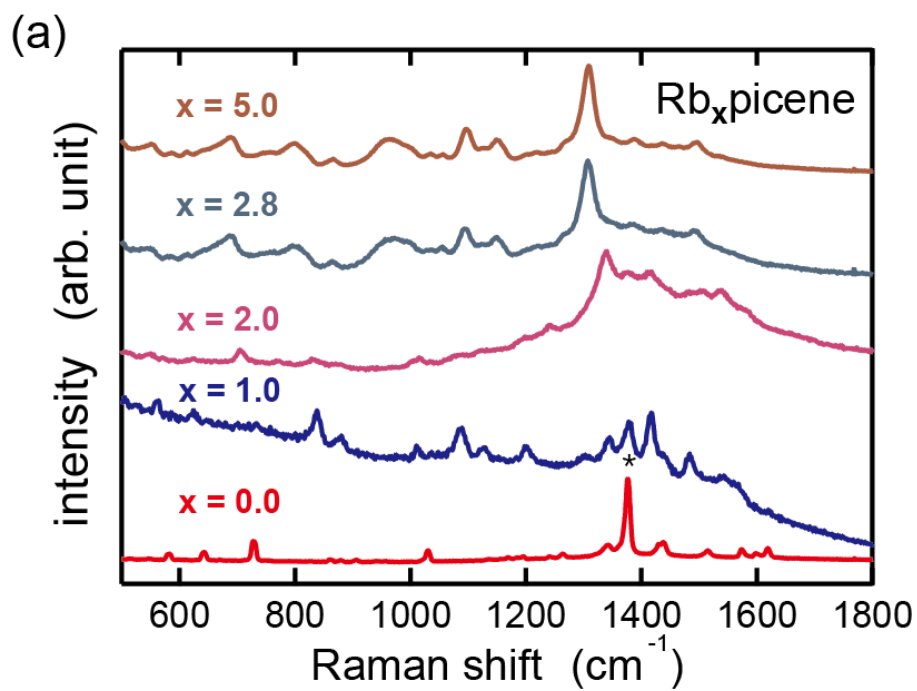

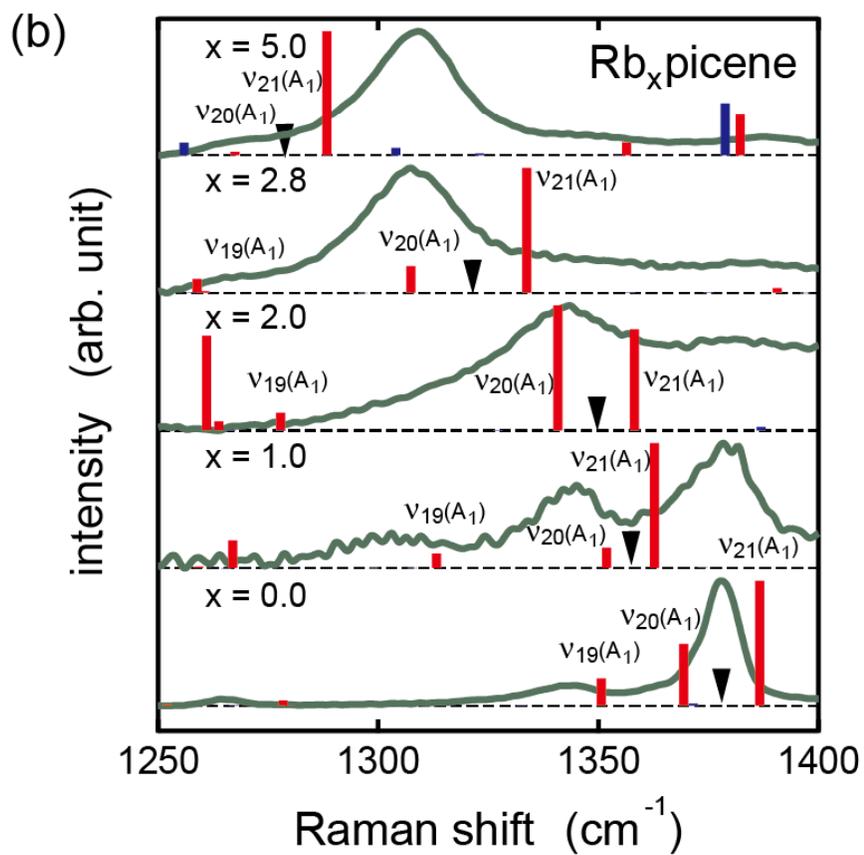

Figure 2. T. Kambe et al.,

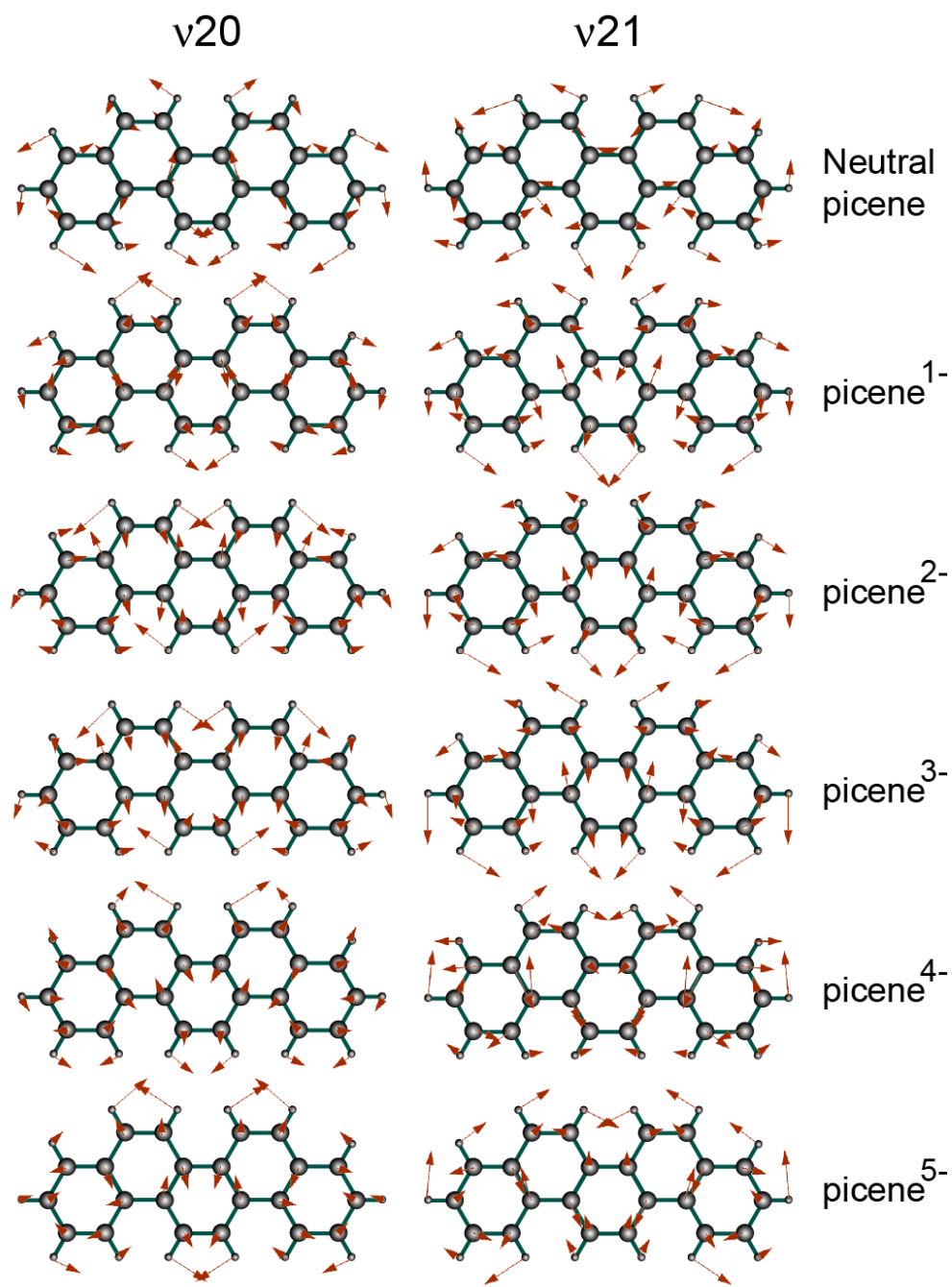

Figure 3. T. Kambe et al.,

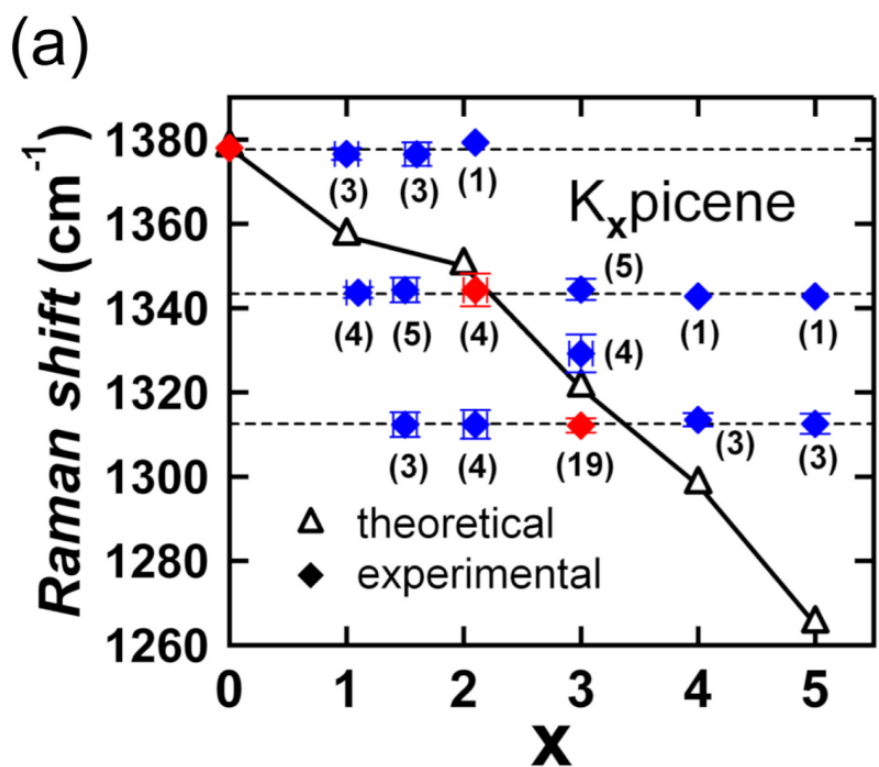

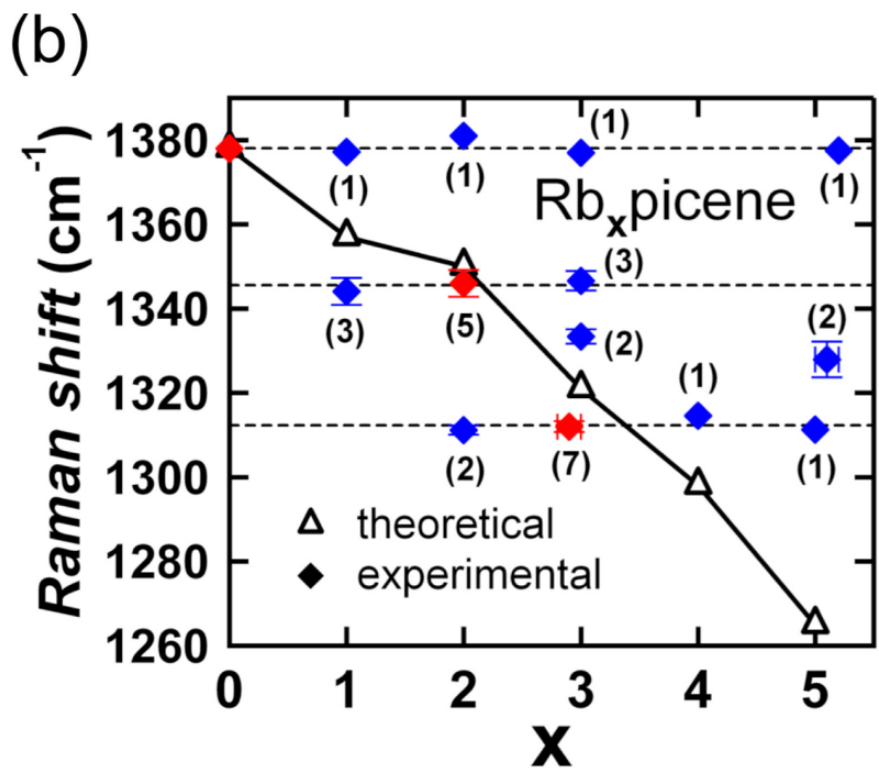

Figure 4. T. Kambe et al.,

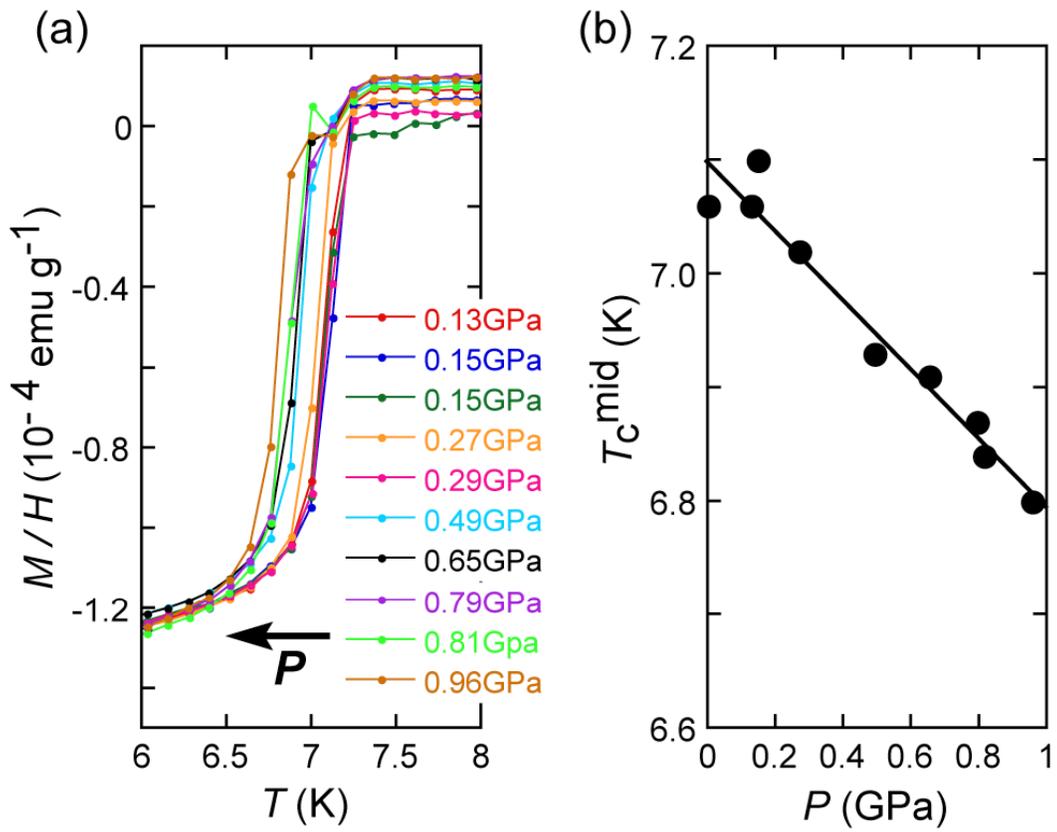

Figure 5. T. Kambe et al.,

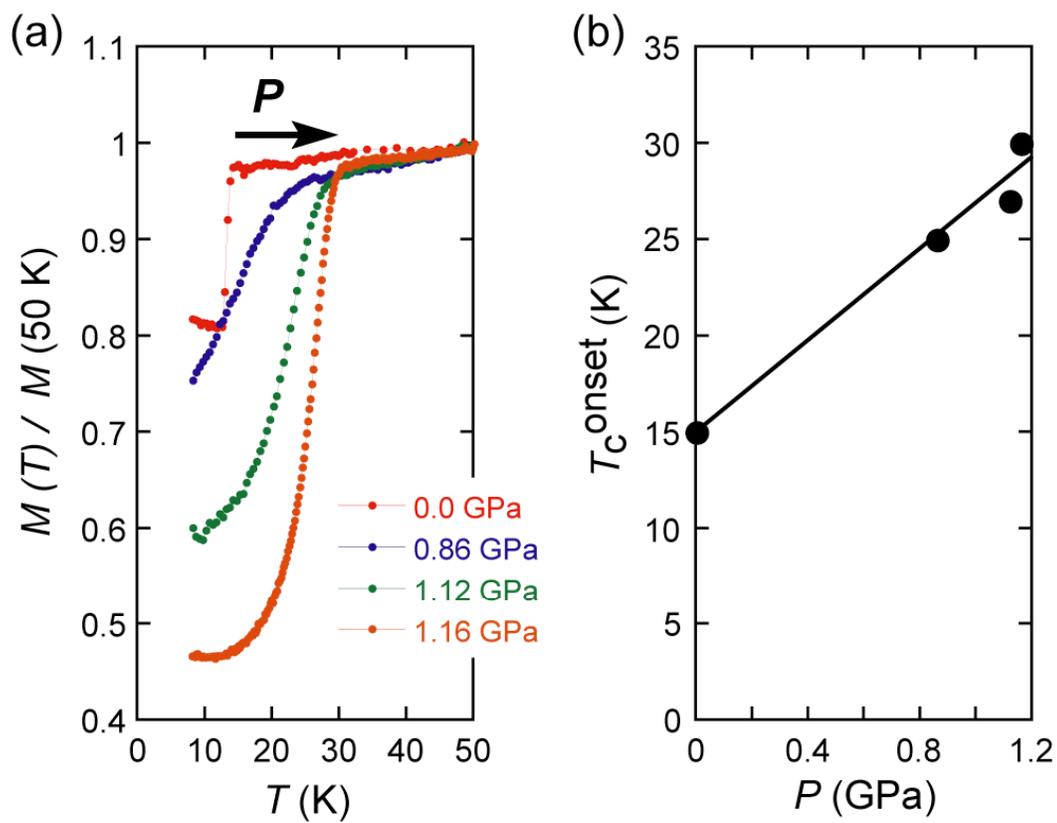

Figure 6. T. Kambe et al.,

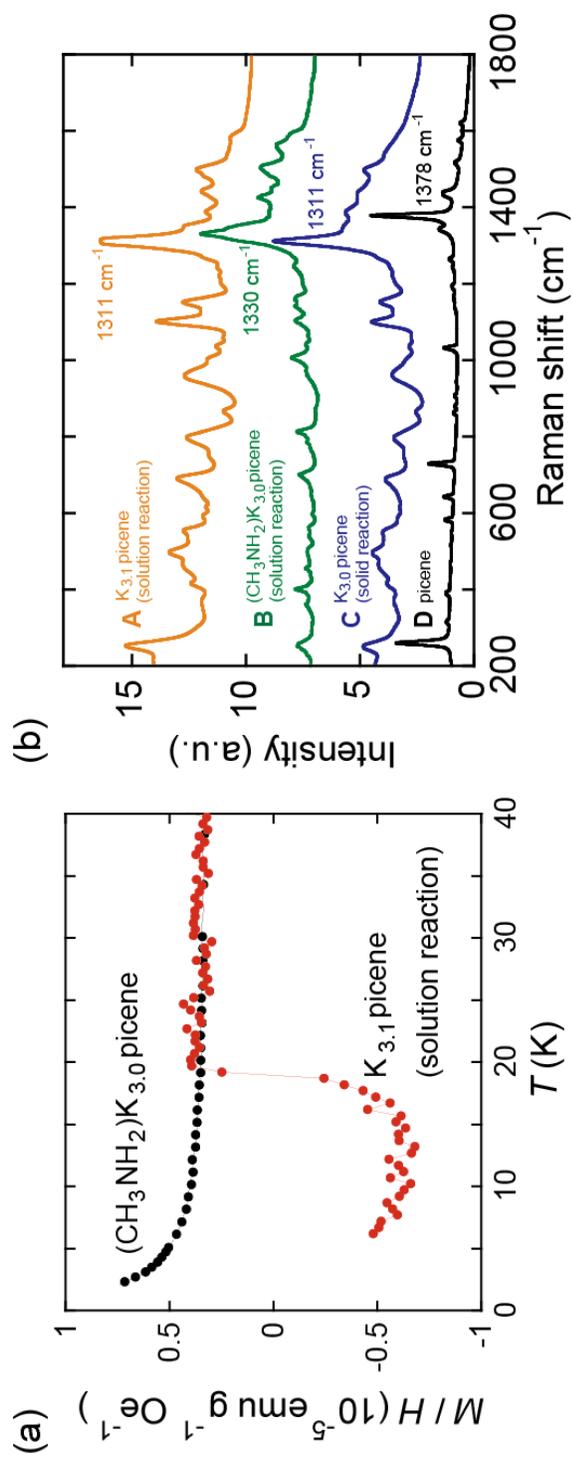

Figure 7. T. Kambe et al.,

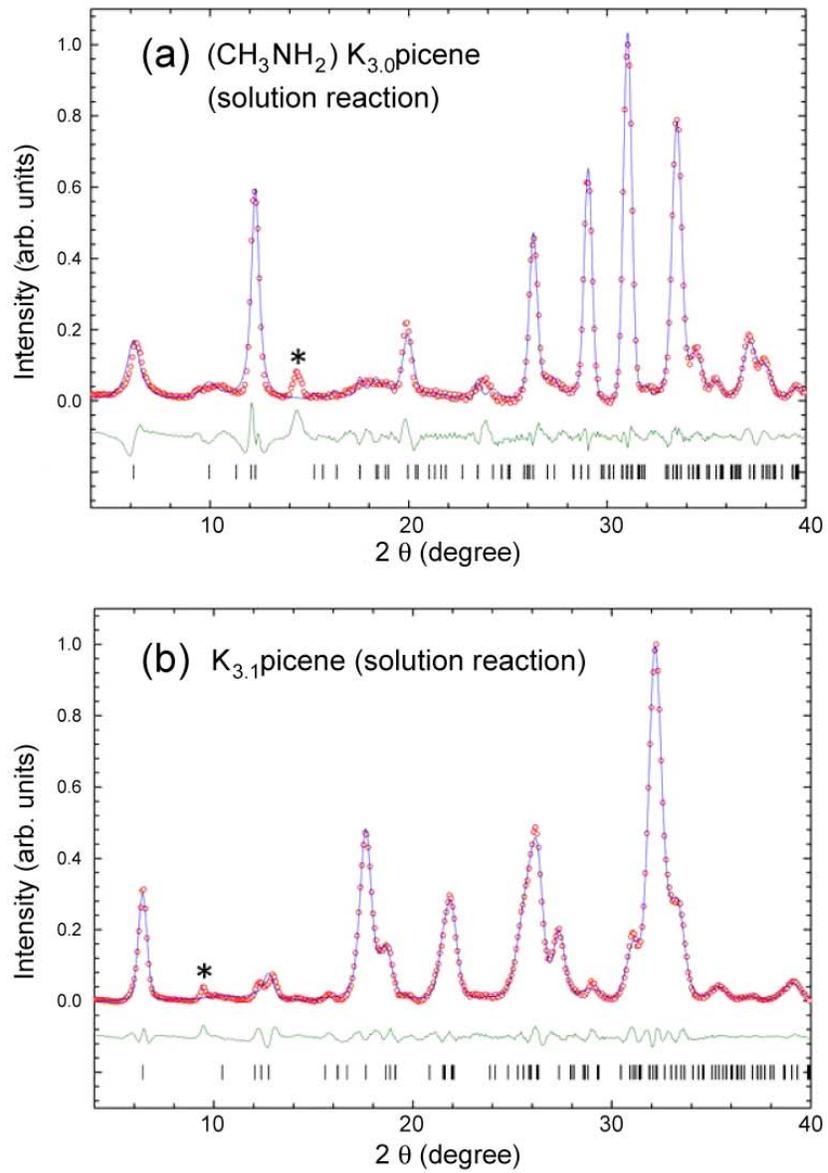

Figure 8. T. Kambe et al.,